\documentclass[aps,prl,amsmath,twocolumn,floats,floatfix,superscriptaddress,nofootinbib,showpacs]{revtex4-1}
\usepackage{amsfonts,amsmath,units,wasysym,epsfig,graphicx,verbatim,color,subfigure,graphicx}
\usepackage{amsmath}
\usepackage{amssymb}
\usepackage{amsfonts}
\usepackage{hyperref}
\usepackage{bm}
\usepackage{color}
\usepackage[normalem]{ulem}
\usepackage{natbib}
\usepackage{graphicx}
\usepackage[utf8]{inputenc}

\usepackage{xcolor}

\def\be{\begin{equation}}
\def\ee{\end{equation}}
\def\bea{\begin{eqnarray}}
\def\eea{\end{eqnarray}}

\newcommand{\ess}{\mathcal{S}}
\newcommand{\ech}{\mathcal{H}}

\newcommand{\qtilde}{\tilde{q}}

\usepackage{natbib}
\usepackage{graphicx}

\newcommand{\IUCAA}{\affiliation{Inter-University Centre for Astronomy and Astrophysics, Post Bag 4, Ganeshkhind, Pune 411 007, India}}
\newcommand{\ICTS}{\affiliation{International Centre for Theoretical Sciences - Tata Institute of Fundamental Research, Survey No. 151, Shivakote, Hesaraghatta Hobli, Bengaluru - 560 089, India.}}

\begin{document}
\title{The Shear at the Common Dynamical Horizon in Binary Black Hole Mergers and its Imprint in their Gravitational Radiation}
\bibliographystyle{unsrt}
\author{Vaishak Prasad}
%\email{vaishak.p@icts.res.in}
\ICTS
\IUCAA

\date{\today}

\begin{abstract}
We study the correlation between a part of the gravitational field at the common dynamical horizon in the strong field regime and the news of the gravitational radiation received from the system in the weak field regime, in the post-merger phase of quasi-circular, non-spinning binary black hole mergers using numerical relativity simulations. We find that, as in the inspiral phase \href{https://journals.aps.org/prl/abstract/10.1103/PhysRevLett.125.121101}{Phys.Rev.Lett.125,121101}, the shear of the common dynamical horizon formed late into the inspiral continues to be well correlated with the news of the outgoing gravitational radiation even at early times. We show by fitting that the shear contains certain quasi-normal frequencies and information about the masses and spins of the remnant and the parent black holes, providing evidence to support the horizon correlation conjecture holds for dynamical horizons in binary black hole mergers. 
\end{abstract}
\maketitle
\section{Introduction}

Gravitational waves from astrophysical systems such as binary black hole mergers provide a new window and means to study and understand the cosmos and particularly the nature of black holes. The detection of gravitational waves from such systems starting with GW150914 \cite{Abbott:2016blz}, 
\cite{LIGOScientific:2018mvr,TheLIGOScientific:2016pea,Nitz:2018imz,Nitz:2019hdf,Venumadhav:2019lyq,Zackay:2019tzo}, \cite{theligoscientificcollaboration2021gwtc21}, \cite{theligoscientificcollaboration2021search} have provided evidence for the existence of black holes and the validity of General Relativity. 

The gravitational radiation received from binary black hole systems travels from faraway regions before reaching us and being recorded on the detectors. Thus we only have access to the weak field regime of such gravitating systems. However, there exist other boundaries of spacetimes that are not directly accessible to us and through which gravitational radiation can leave the system. These exist as the boundaries of the black hole regions in the strong field regimes of such binary black hole systems. Probing the nature of gravity in this region of spacetime will provide us with new and important information about strong gravitational effects, black holes, the validity of the no-hair conjecture and General Relativity, cosmic censorship, etc.

The gravitational radiation being recorded in the detectors can be used to infer the parameters of the system, such as their masses, spins, location and spatial orientation, etc. However, it would be very useful to know if these waves signifying the weak field dynamics carry any additional information about the strong field gravitational effects of the system. This would allow one to infer and study the strong field dynamics of the BBH system using the gravitational radiation received in faraway regions and can be used to probe the strong gravity physics of black holes and general relativity. One such prospect is presented by what we call the horizon correlation conjecture, which states that in a binary black hole merger scenario, the strong field and weak field gravitational fields are correlated.

In the literature, many authors have suspected such a correlation to exist in the case of a single perturbed black hole. In such a system, the source of gravitational radiation could be intuitively pictured to originate from a region approximately localized around the peak of the gravitational potential, which would then scatter and disperse to other regions of the spacetime as the system evolves. While most of this radiation escapes out to future null infinity, a part of it travels inwards towards the inner boundaries of spacetime and is absorbed at the horizons of the black hole. One thus expects that are escaping the system through these boundaries may be correlated due to the common origin or source of these these waves. Based on such an intuitive picture, several authors have attempted to study and find evidence for the conjecture. In particular, \cite{Jaramillo:2012rr} develops some analytical techniques and ideas to study the conjecture and \cite{Gupta:2018znn} suspected it to hold in the post-merger phase of BBH mergers.

The first systematic attempt to study this conjecture quantitatively was made in \cite{shear-news2020}, where we also attempted to study the validity of the conjecture in the inspiral phase of binary black hole mergers. We proposed a new framework and approach to investigate the conjecture using the notion of dynamical horizons, numerical relativity and elements of data analysis.
In particular, we use the notion of dynamical horizons to describe the surfaces of black holes and propose to study the correlations between the gravitational fields at the dynamical horizons and future null infinity using Numerical Relativity. 

Isolated and Dynamical horizons are three-dimensional null and space-like surfaces respectively that describe the boundaries of the black hole regions. They have several advantages over event horizons. First, they are free from the latter's teleological nature. Event horizons are global in nature and require the notion of a complete conformal future null infinity \cite{GerochIncomplScriPlus}. One would need complete information about the spacetime future and the past in order to locate the event horizons. Due to these restrictions, event horizons are known to be insensitive to the local physics of the system. e.g., event horizons can form and grow in empty regions of spacetime in anticipation of matter/energy that would fall in. They can also be non-smooth, have creases \cite{gadioux2023creases} and no preferred way to foliate them. On the other hand, dynamical horizons are smooth and quasi-local. Several important results originally associated with Killing horizons, like the laws of black hole mechanics, have been extended to Dynamical horizons \cite{Ashtekar:2004cn} and thus are physical in nature. These dynamical horizons lie behind event horizons and are thus causally disconnected from outside observers.

For the benefit of the reader, we summarize the results of the previous work \cite{shear-news2020} here. We ran a small set of numerical simulations of non-spinning binary black holes of varying mass ratios on quasi-circular orbits. We track their dynamical horizons through the inspiral, merger and ringdown and compute various quasi-local quantities on them. In the inspiral phase, there are two dynamical horizons, one for each black hole. We then compute the shear of each of the dynamical horizons and study their evolution through the inspiral. The shear of the dynamical horizon is a quasi-local, spin-2 field that quantifies a part of the tidal coupling between the black holes, and the amount of transverse gravitational radiation infalling at the respective dynamical horizons. In \cite{shear-news2020}, it was interestingly found that the shear of each of the dynamical horizons has a familiar chirp-like behaviour. Additionally, the shears were shown to be strongly correlated with the news of the outgoing gravitational radiation from the system. It was also found that the shears can be described by simple phenomenological relations to the news, thus allowing one to compute them approximately at the dynamical horizons of the black holes in the strong field regime using only the gravitational radiation received at faraway regions. To further interest, we showed that the parameters of the black holes can be fully recovered using these shears as signals against a bank of templates of gravitational news. This was the first time that evidence for the horizon correlation conjecture was found to hold in the inspiral phase of the BBH system using numerical relativity.

In this study, we extend the same analysis to the post-merger phase. In particular, we study the correlation between the shear of the common horizon of the system that forms close to the merger and the News of the gravitational waves recorded by a detector at a large distance from the black hole centre. The post-merger dynamics of black holes resulting from the numerical simulation of the due process of a merger is non-trivial. Although it is fairly easy to evolve binary black hole systems since the first breakthroughs \cite{Pretorius:2005gq, Pretorius_2005, Campanelli:2005dd, Baker:2005vv, Campanelli2006lo} and extract gravitational radiation, the limited computational resources can restrict the resolution of the runs and affect the long term accuracy and stability of the simulations. As we also show here, the strong field dynamics, especially at the dynamical horizons of black holes, can accumulate some inaccuracies from the long-term evolution of the system, and the dynamical horizon of the final black hole need not be exactly Kerr but only up to some accuracy level.

\subsection{Numerical Simulations}
\begin{table}
\begin{tabular}{|p{2cm}|p{2cm}|p{2cm}|p{2cm}|}
 \hline
 $q $ & $D/M$ & $p_r/M$& $p_{\phi}/M$\\
 \hline
0.85    &   12.0    & -0.000529     & 0.08448\\
0.75    &   11.0    & -0.000686     & 0.08828\\
0.667   &   11.75   & -0.000529     & 0.08281\\
0.5     &   11.0    & -0.000572     & 0.0802\\
0.25    &   11.0    & -0.000308     & 0.05794\\
0.1667  &   10.5    & -0.000219     & 0.04590\\ 
 \hline
\end{tabular}
\caption{Initial parameters for non-spinning binary black holes with quasi-circular orbits. $q=M_2/M_1 < 1$ is the mass ratio, $D$ is the initial separation between the two holes, and $p_r$ and $p_{\phi}$ are the initial radial and azimuthal linear momenta of the punctures respectively.}
\label{tab:ic_qc0}
\end{table}
The system setup is identical to that in \cite{shear-news2020}, consisting of a non-spinning binary black hole system in a quasi-circular configuration. The simulations are 
performed using the publicly available Einstein Toolkit framework 
\cite{Loffler:2011ay, EinsteinToolkit:web}. The initial data is generated 
based on the puncture approach \cite{PhysRevLett.78.3606,Ansorg:2004ds}, 
which is then evolved through  BSSNOK formulation
\cite{Alcubierre:2000xu, Alcubierre:2002kk, Brown:2008sb} using the
$1+\log$ slicing and $\Gamma$-driver shift conditions.  Gravitational 
waveforms are extracted \cite{Baker:2002qf} on coordinate spheres at 
various radii between $100M$ to $500M$.  
The computational grid set-up is based on the multi-patch approach using 
Llama \cite{PhysRevD.83.044045} and Carpet modules, along with adaptive 
mesh refinement (AMR). The various horizons are located using the method 
described in \cite{Thornburg:1995cp, Thornburg:2003sf}. 
General quasi-local physical quantities are computed on the horizons following 
\cite{Dreyer:2002mx, Schnetter:2006yt}. Several post-processing tools to carry out analysis of numerical relativity data have been developed in Python \cite{wftools} and used here. 

We consider non-spinning binary black hole systems with varying mass ratios 
$q=M_2/M_1$, where $M_{1,2}$ are the component horizon masses ($M_1\geq M_2$). 
We use the GW150914 parameter file available from \cite{wardell_barry_2016_155394} as
a template. For each of the simulations, as input parameters we provide initial
separation between the two punctures $D$, mass ratio $q$ and the radial and azimuthal linear momenta $p_r$, $p_{\phi}$ respectively, while keeping the total physical horizon masses $M=M_1+M_2=1$ fixed in our units. Parameters are listed in Tab.~\ref{tab:ic_qc0}. We compute the corresponding initial locations, the $x$, $y$, and $z$ components of linear momentum for both black holes and grid refinement levels, etc., before generating the initial data and evolving it. 

For computing the quasi-local quantities, we use a uniform angular grid of size ($36, 74$) on each dynamical horizon. The spins and masses of the black holes were computed using the quasi-local computations.
We also carry out a convergence test of the horizon data by simulating at two different grid resolutions of the horizon. Our simulations agree very well with the catalogue simulations \cite{RITcatalog:web}, with merger time discrepancies of less than a few per cent.  

\section{Basic notions}

A dynamical horizon $\ech$ is a space-like three-surface that is foliated by a sequence of closed two-dimensional space-like surfaces $\ess$ of spherical topology. Adopting the Newman-Penrose formalism with complex null basis vectors $(n^{\pm}), m$, these surfaces have zero outward expansion and are thus marginally trapped:
\begin{equation}
    \Theta_{+} = \qtilde^{ab} \nabla_a n^{+}_b = 0 \label{shear}
\end{equation}
In this work, we are interested in the shear of the outward normal of these surfaces $\ess$. The shear of a dynamical horizon is a quasi-local tonsorial quantity and can be described as a spin-2 tensorial field on each slice of the dynamical horizon. Alternatively, it can be described as a complex scalar in the spin-coefficient formalism. It has two independent components in the spin-coefficient formalism and can be defined as:
\begin{equation}
    \sigma = m^a m^b \nabla_{a} n^+ _{b}
\end{equation}
Due to the zero expansion and being normal to the MTS, it is transverse and traceless and quantifies the transverse gravitational radiation infalling at the horizon. The shear is a dominant quantity appearing in the balance law at the dynamical horizon and forms an important part of the tidal coupling between the black holes in the strong field regime \cite{shear-news2020, Ashtekar:2000sz}.

We track the common dynamical horizon through the post-merger phase for all the simulations listed in Tab.~\ref{tab:ic_qc0} immediately after its formation and compute the shears as defined in \eqref{shear}.

In this study, we attempt to find a correlation between this quantity i.e. the shear of the common dynamical horizon, and the news of the gravitational radiation emitted by the system, computed using the Weyl tensor component $\Psi_4$ extracted on spheres of large radii (typically $100 - 500M$):
\begin{equation}
 \Psi_4 = C_{abcd}n^a\bar{m}^bn^c\bar{m}^d\,.
\end{equation}

The news is defined as: 
\begin{equation}
  \mathcal{N}^{(\ell,m)}(u) = \mathcal{N}^{(\ell,m)}_+ + i\mathcal{N}^{(\ell,m)}_\times = \int_{-\infty}^u\Psi_4^{(\ell,m)}\,du\,.
\end{equation}

We refer the reader to \cite{shear-news2020} for further details.

A common dynamical horizon formed from the merger of the individual dynamical horizons late into the inspiral (henceforth the common horizon) and is also foliated by marginally outer trapped surface (MOTS). When formed, as shown here, it is in a highly deformed state, far from symmetry. 

To facilitate the study and for convenience, we make a few approximations. Firstly, as mentioned above, the remnant black hole formed from the merger of individual holes is spinning and highly distorted. When studying gravitational radiation from such a system, it is often recognized that overtones and non-linearities would be present in the signal right after the common horizon is formed, and the analysis is begun around $10M$ after the peak of the signal. Here, owing to the resolution of our runs, we do not start later but at the peak of the signal. The study of overtones and non-linearities would require much higher resolution runs along with higher sampling in time, which we reserve for a later study. The purpose of this work is to study the correlations between the physics of the dynamical horizons and future null infinity in the post-merger phase. Secondly, the shear of a dynamical horizon is a quasi-local tonsorial quantity and can be described by as a spin-2 field on each slice of the dynamical horizon. Alternatively, it can be described as a complex scalar field in the spin-coefficient formalism. It has two independent components in the spin-coefficient formalism at all points on the dynamical horizon. For the purposes of this study, we only use the shear computed at one of the equatorial points of the slices of the dynamical horizon. 

The common horizon eventually settles down to the outer horizon of the remnant black hole asymptoting to the Kerr isolated horizon, by absorbing the appropriate amount of radiation quantified majorly by the shear and gaining symmetry in the process.

\section{Methodology}

In order to study the correlations between the weak and strong field dynamics, we follow a simple, multi-pronged approach. Treating the shear of the common dynamical horizon as the signal, we carry out the following tests:
\begin{enumerate}
    \item Remnant parameter estimation. We fit the shear to free i.e. unconstrained damped sinusoids and compare it with the QNM frequency of the final Kerr black hole. Assuming the mass, we also attempt to infer the spin of the remnant black hole from the best-fit frequencies. 
    \item Parent parameter estimation. We use the parameter estimation routine to carry out the parameter estimation of post-merger numerical relativity data using elements of gravitational wave data analysis, which is also parallelized. It was originally developed and used first in \cite{shear-news2020}. The parameter estimation routine was used to infer the mass ratio and the chirp mass of the parent black holes of the binary system from the information in the strong field dynamics at the dynamical horizons of BBH mergers. We optimize a least squares figure of merit from a template bank of waveforms using the waveform approximant SEOBNRv4PHM. 
\end{enumerate}

In the next two sections, we present results that strongly suggest that the shear-news relationship that conveys the strong correlation between strong field and weak field dynamics is valid in all phases of binary black hole mergers.

\section{Results}
\begin{figure*}
\includegraphics[width=\columnwidth]{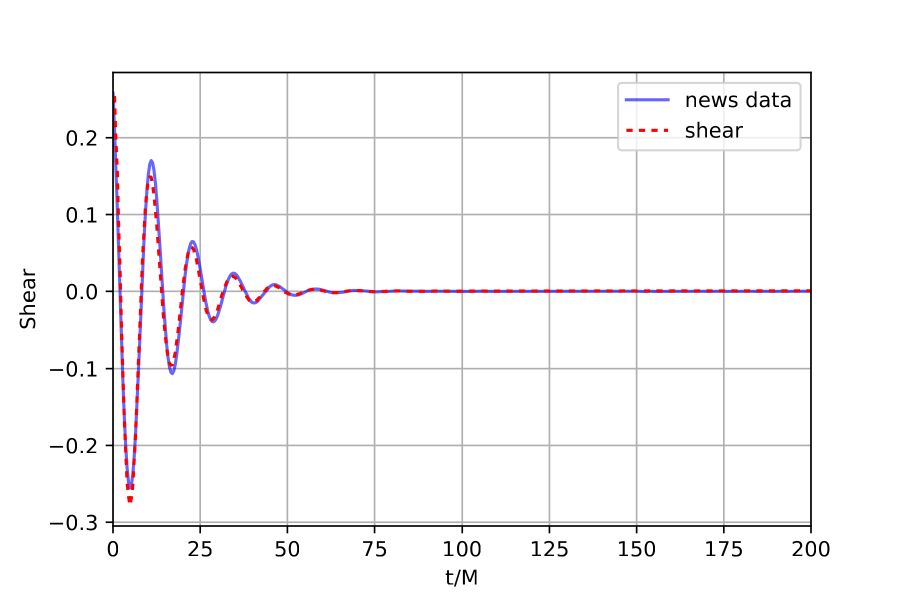}
\includegraphics[width=\columnwidth]{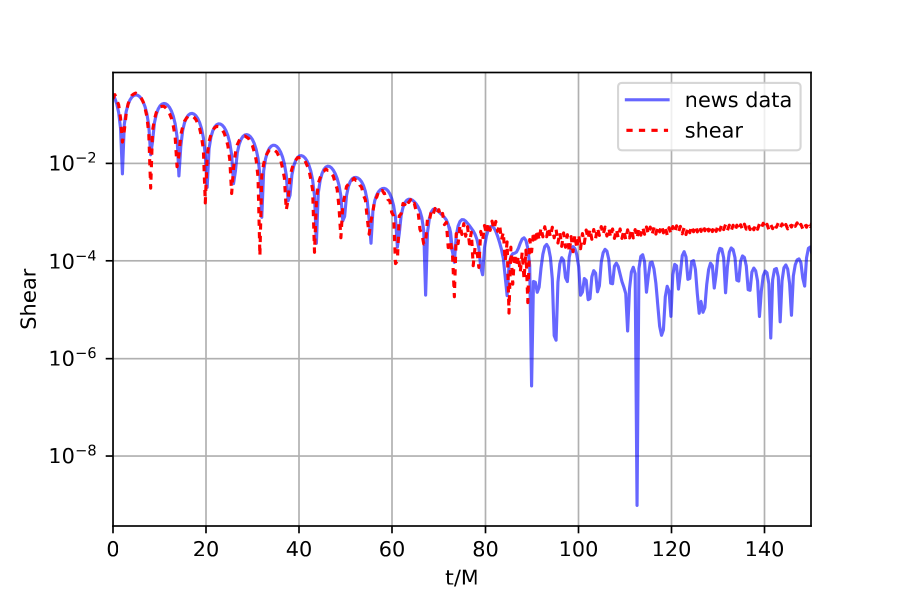}
\caption{The shears of the common horizon aligned with the numerical news computed from the $\Psi_4$ for the simulation $q=0.6$. In the left panel, the real part of the shear is shown, and on the right, the same data is shown on a logarithmic scale on the y-axis.}
\label{fig:shear_news}
\end{figure*}

\subsection{Shear-News match}

How well is the News of the outgoing gravitational radiation correlated with the shear of the dynamical horizon? To answer this, we align the shears with the news computed from the extracted Weyl scalar component $\Psi_4$ of the numerical simulation, suitably in phase and time by maximizing the least squares figure of merit. A convenient observation is that when aligned in time, the time shift between the shear and the news is consistent with the extraction radii of the spheres on which the latter is extracted in a numerical domain. 

When aligned in time and phase across the simulations, we derive a phenomenological relation that relates the shear of the common dynamical horizon and the News of the outgoing gravitational radiation across the simulations:
\begin{equation}
    \left\lvert \dfrac{\sigma}{\mathcal{N}} \right \vert = 24.81 q^2 -30.64 q +  10.62
\end{equation}
These coefficients are roughly the same for both the components of the shears and we conjecture them to be universal. This is an interesting parallel with the results for the inspiral phase as presented in Eq.6 of \cite{shear-news2020}. Fig.~\ref{fig:shear_news} shows the match between the two signals graphically. 
Thus they are very strongly correlated.

\subsection{Fits to damped sinusoids}
\begin{figure}
\includegraphics[width=\columnwidth]{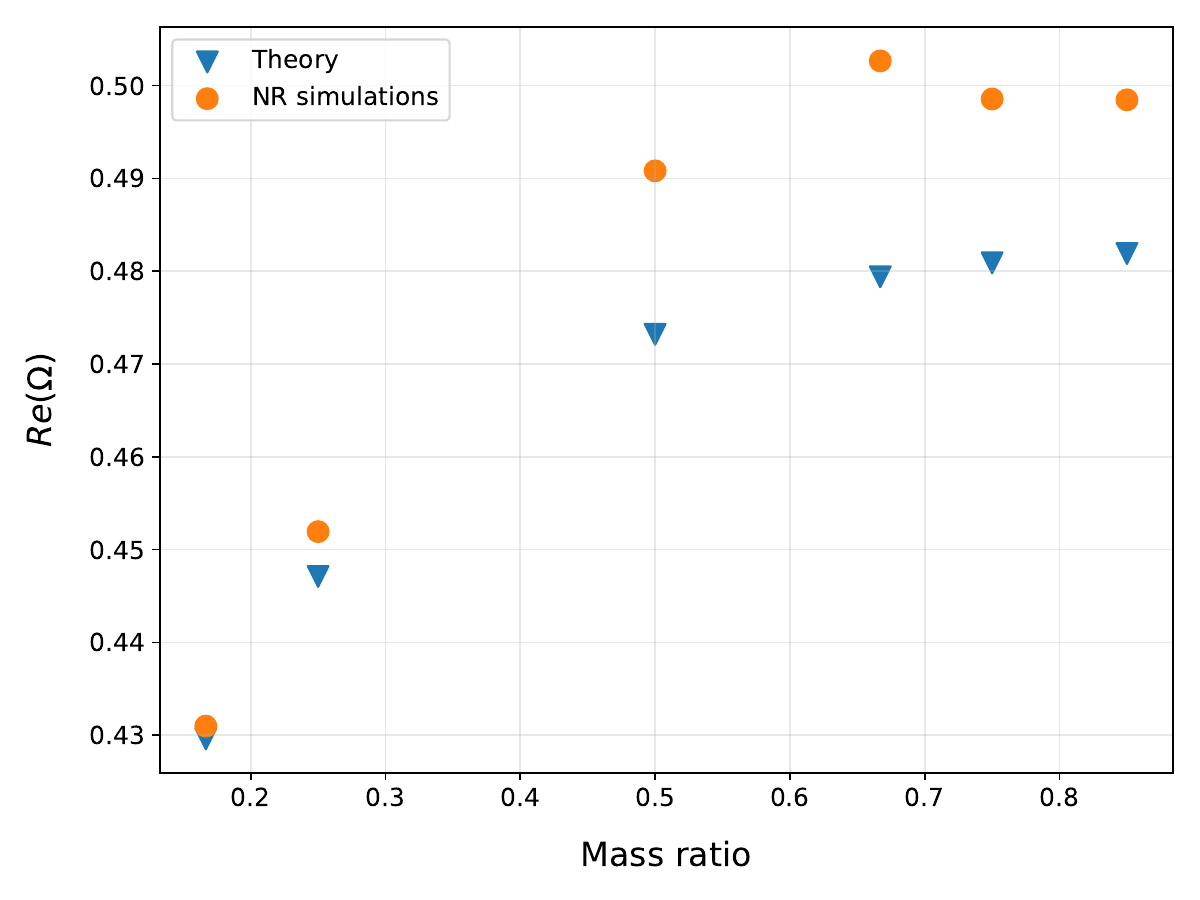}
\caption{A plot of the real part of the fundamental QNM frequency obtained by fitting the shear of the common dynamical horizon. The corresponding values from perturbation theory are shown as blue triangles. The numerical values are displayed in Tabs.~\ref{tab:actual_qnms} and \ref{tab:shear_qnms}.}
\label{fig:re_omega}
\end{figure}
\begin{figure}
\includegraphics[width=\columnwidth]{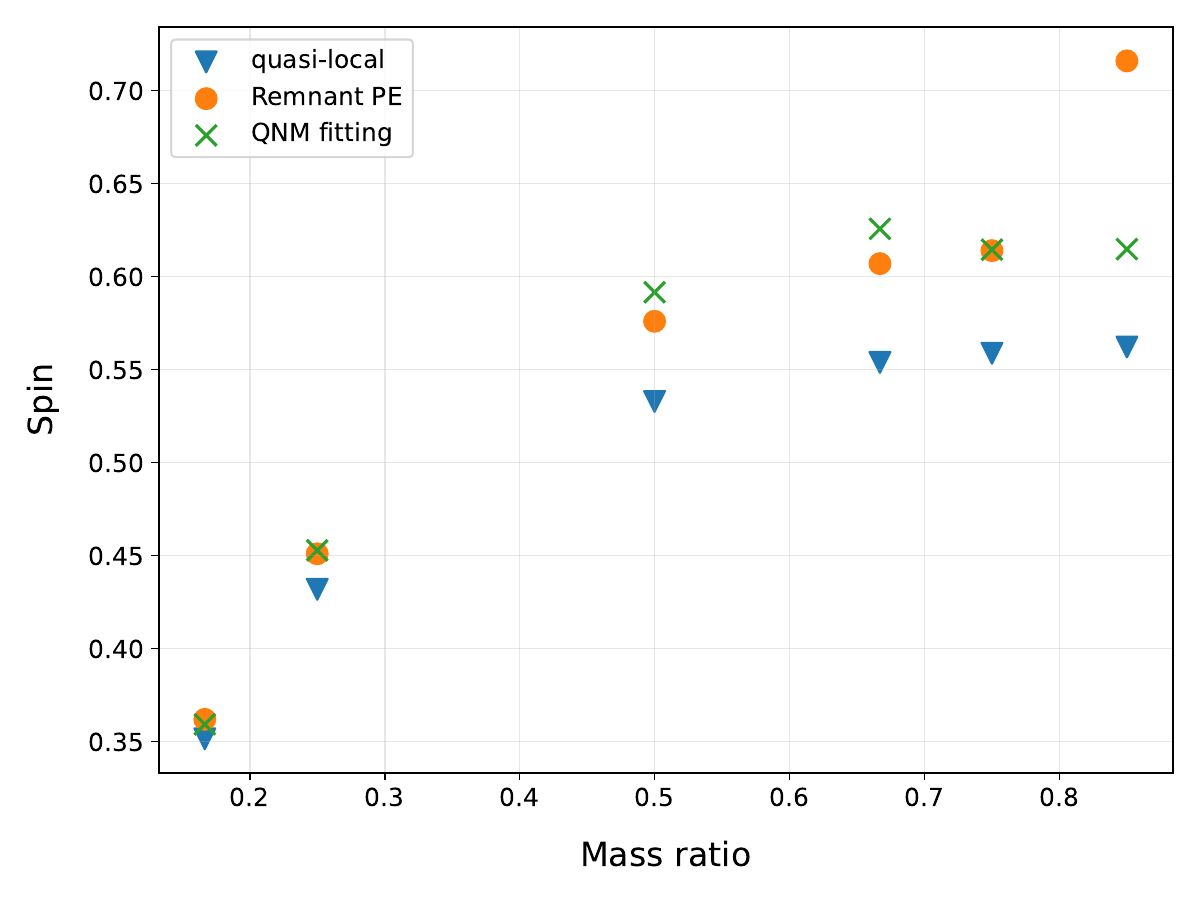}
\caption{Recovery of the spin of the final black hole. The value of the spin of the black hole estimated from the fit to the unconstrained model is shown as green crosses. The estimates from the PE routine is shown in orange circles. The actual values obtained from quasi-local computations is shown as blue inverted triangles. The numerical values of these spin estimates for the latter two are displayed in column 5 of Tab.~\ref{tab:shear_remnant_pe} column 2 of Tab.~\ref{tab:actual_qnms}) respectively.}
\label{fig:shear_spin}
\end{figure}

Does the shear of the common dynamical horizon have Quasi-normal behaviour? Can the spin of the remnant black hole be estimated using the shear of the common dynamical horizon?

In order to answer these questions without assuming general relativity, we attempt to estimate the final spin of the remnant from the shear of the common horizon by fitting it to free, unconstrained damped sinusoids.  

From black hole perturbation theory, we know that the waveform past the temporal location of maximum intensity can be well approximated by a superposition of quasi-normal modes of the final black hole. Owing to the no-hair theorem, each QNM frequency corresponds to a unique set of physical parameters of the final black hole (Mass, Spin). A remnant black hole's linear perturbations can be expressed as a superposition of quasi-normal modes. The angular behaviour can be decomposed in terms of spin-weighted spheroidal harmonics and each quasi-normal mode can be labelled by the indices $(n, \ell, m)$, for each spin weight $s$. Two different modes will in general have different Amplitudes,  frequencies $\omega$ and damping rates $\gamma$. 

The temporal behaviour of these functions ( at a particular coordinate location) can be written down as:
\begin{equation}
    f(t) = \Sigma_{n, \ell, m} A_{n\ell m} e^{(-\gamma_{n\ell m} t)} \sin{(\omega t + \phi)}
\end{equation}
In this exercise, we adopt a fully agnostic view of the questions posed above and approach using an unconstrained fitting model. Each mode $(n, \ell, m)$ will add five free parameters in an unconstrained fitting model. Thus the unconstrained fits are challenging from a computational and data analysis point of view. 

The gravitational News can be obtained through a single derivative of the strain (or an integral of $\Psi_4$) and is therefore also quasi-normal in behaviour. Therefore, if the shear-news correlation holds in the post-merger phase, one may expect the same quasi-normal behaviour for the shear of the common outer horizon as with the gravitation News (and therefore the strain) of the system. 

 The shear of the common horizon \eqref{shear} was fit to two damped sinusoids. We could recover the ($n=1, \ell=2, m=2$ ) mode in most of the cases, with limited accuracy. The frequency of the recovered fundamental mode is also shown in Fgg.~\ref{fig:re_omega}. The details of the recovered fundamental QNM frequency are listed in Tab.~\ref{tab:shear_qnms}. A table containing the QNM details from analytic computations is shown in Tab.~\ref{tab:actual_qnms} for reference and our estimates can be seen to agree well with these. 
 
 Using the fundamental QNM mode and the final mass of the remnant as reported by the numerical simulations, the spin of the remnant was computed. This was found to be close to the final spin as computed using quasi-local techniques from the numerical simulation. The spin of the remnant was computed from the numerical simulation using the $\ell=1$ spin multipole moment of the common horizon in the quasi-local formalism. The details are represented in Figs.~\ref{fig:shear_spin}. Here, we assume that the final spin of the black hole formed is aligned with the $\hat{z}$ axis (which is confirmed by the numerical data). In these figures, one can visually see a Price-like tail in the late time evolution of the shear. However, the Price power law could not be confirmed owing to the poor resolution of data.

\begin{table}
\begin{tabular}{|p{1.8cm}|p{1.8cm}|p{1.8cm}|p{1.8cm}|p{1.8cm}|p{1.8cm}|}
\hline
Simulation  &   Spin        &   $\omega_{(0)}$   &   $\gamma_{(0)}$      \\
\hline
0.85     &   5.622e-01   &   4.819e-01       &   8.457e-02           \\
0.75     &   5.588e-01   &   4.809e-01       &   8.463e-02           \\
0.6667   &   5.539e-01   &   4.794e-01       &   8.473e-02           \\
0.5      &   5.329e-01   &   4.732e-01       &   8.511e-02           \\
0.25     &   4.319e-01   &   4.471e-01       &   8.654e-02           \\
0.1667   &   3.515e-01   &   4.296e-01       &   8.733e-02           \\
\hline
\end{tabular}
\caption{Theoretical QNM data for a black hole of mass 1 from perturbation theory.}
\label{tab:actual_qnms}
\end{table}
\begin{table*}
\begin{tabular}{|p{1.8cm}|p{1.8cm}|p{1.8cm}|p{1.8cm}|p{1.8cm}|p{1.8cm}|}
\hline
Simulation  &   $\omega_{(0)}$   &   std. error    &   $\gamma_{(0)}$  &   std. error     & Amplitude \\
\hline
0.85        &   5.236e-01       &   8.354e-04   &   7.805e-02       &   7.662e-04     &  0.224 \\
0.75        &   5.226e-01       &   4.245e-04   &   8.076e-02       &   4.879e-04     & 0.232 \\
0.6667      &   5.258e-01       &   1.088e-03   &   8.263e-02       &   9.433e-04     & 0.232 \\
0.5         &   5.102e-01       &   3.510e-03   &   8.055e-02       &   3.834e-03     & 0.215 \\
0.25        &   4.621e-01       &   2.840e-03   &   8.490e-02       &   2.512e-03     & 0.07  \\
0.1667    &   4.371e-01       &   6.928e-04   &   8.383e-02       &   7.394e-04     & 0.043 \\
\hline
\end{tabular}
\caption{The details of the fundamental quasi-normal frequencies that best fit the shear $\sigma$ of the common horizon of the system and the associated standard fitting errors.}
\label{tab:shear_qnms}
\end{table*}

We also followed another approach to estimate both the masses and the spins of the remnant from the shear of the common outer horizon, using a procedure that minimizes a least squares figure of merit over the dominant/ lowest QNM. The recovered parameter values are listed in Tab.~\ref{tab:shear_remnant_pe} and the spin parameter values are shown in Fig.~\ref{fig:shear_spin}.

One overtone, corresponding to the modes ($n=1, \ell=2, m=+ 2$) for $q=0.85$ with $9.5\%$ deviation and ($n=1, \ell=2, m=-2$) for $q=0.75$ could be estimated. The difficulty in the estimation of the overtones can be associated with the limited resolution of the numerical simulations. A good estimate of the overtones would require a much higher sampling rate of the data output along with the higher resolution of the run. A significant improvement of the results presented here can be expected from higher-resolution simulations which we intend to investigate later. There are a few features of the fits which we wish to mention here.

It was observed that adding an additional damped sinusoid improved the estimation of the fundamental mode. However, the additional mode could not always be identified with an overtone. The physical nature of these spurious modes (over and above the fundamental mode) is unknown, and in our opinion is associated with numerical noise/ inaccuracies in the data. Thus, although the waveforms may agree with the expected QNM results, the quantities on the dynamical horizons will have larger errors accumulating from the long-term evolution of the system making them excellent probes of the measure of the quality of the numerical simulations. 

At late times, the dynamical horizon should asymptote to the Kerr isolated horizon on which the shear vanishes. In our case, the shears reach a numerical noise floor of $10^{-3}$, which is consistent with the errors on the QNM frequencies estimated here. Further work and computing resources is needed to verify whether the noise floor improves with the resolution of the runs.

\subsection{Remnant Parameter Estimation}

\begin{table*}
\begin{tabular}{|p{2cm}|p{2cm}|p{2cm}|p{2cm}|p{2cm}|p{2cm}|p{2cm}|p{2cm}|}
\hline
$\widehat{q} $ & Best matching mass & Actual mass& \% Error &Best matching spin& Actual spin & \% error&Match\\
\hline
0.85    &   0.973   &   0.952   &   2.132     & 0.716    &   0.620   &   15.600  &   0.999   \\ 
0.75    &   0.950   &   0.954   &  -0.349     & 0.639    &   0.614   &   4.009   &   0.996   \\  
0.6667  &   0.955   &   0.956   &  -0.032     & 0.634    &   0.607   &   4.435   &   0.998   \\
0.5     &   0.962   &   0.962   &   0.009     & 0.621    &   0.576   &   7.736   &   0.999   \\ 
0.25    &   0.997   &   0.978   &   1.907     & 0.475    &   0.451   &   5.274   &   0.989   \\ 
0.1667  &   0.992   &   0.986   &   0.64      & 0.3624   &  0.362    &   10.43   &   0.986   \\
\hline
\end{tabular}
\caption{Remnant parameter estimation. The results of parameter estimation using the shear $\sigma$ of the remnant black hole were carried out using the match routine.}
\label{tab:shear_remnant_pe}
\end{table*}

We carry out parameter estimation using the shear and the ringdown-only portion of the templates constructed from the SEOBNRv4PHM family of templates. As discussed earlier, we begin at the peak of the waveform amplitude. We use our routine specifically developed to carry out data analysis with numerical relativity data for this purpose. We carry out the parameter estimation routine with both the final remnant mass and the spin as free parameters. The results are shown in Tab. ~\ref{tab:shear_parent_pe}. 

\subsection{Parent parameter estimation}
\begin{table}
\begin{tabular}{|p{1cm}|p{1cm}|p{1cm}|p{1cm}|p{1cm}|}
\hline
$q$ & $\widehat{q}$ & $\mathcal{M}$ & $\widehat{\mathcal{M}}$ \\
\hline
0.85    & 0.783 & 0.434 &   0.431   \\ 
0.75    & 0.676 & 0.430 &   0.425   \\ 
0.6667  & 0.621 & 0.425 &   0.416   \\
0.5     & 0.500 & 0.401 &   0.401   \\
0.25    & 0.310 & 0.333 &   0.365   \\
0.1667  & 0.184 & 0.284 &   0.308   \\
\hline
\end{tabular}
\caption{Parent parameter estimation. The values of the mass ratio $\widehat{q}$ and chirp mass $\widehat{\mathcal{M}}$ of the best template of the SEOBNRv4PHM family that matches the shear of the common horizon is shown along with those of the punctures $q, \mathcal{M}$ of in initial data.}
\label{tab:shear_parent_pe}
\end{table}

Here, we attempt to estimate the parameters of the parent black holes using the post-merger-only shear data. For this purpose, we carry out a similar parameter estimation procedure as in the previous section. Here, we attempt to find the best matching waveform that minimizes the least squares figure of merit from a pool of templates in the parameter space ($q, \mathcal{M}$) that maximizes match over phase and time generated using the popular waveform approximant SEOBNRv4PHM. 

Since the initial black holes were not spinning in our setup, we only varied the individual masses of the black holes to populate the template space and the spins were fixed to zero. The results are shown in Tab.~\ref{tab:shear_parent_pe}

\flushbottom
\section{Discussion}

In this work, we analyze the shear of the common outer dynamical horizon formed from the merger of two black holes using numerical relativity simulations. In particular, we explore the validity of the horizon correlation conjecture that the strong field and weak field dynamics are strongly correlated. For this purpose, we analyze a set of 6 non-spinning binary quasi-circular black hole mergers. For the analysis, we use two quantities: the shear of the common outer horizon of the system $\sigma$ and the News of the system  the post-merger phase, recorded in a detector stationed at a faraway distance of $r = 100M$ units from the centre of the system.  

%In order to test the validity of the horizon correlation conjecture, we explored multiple scenarios and multiple pieces of evidence for the correlations. Firstly, we showed that the shear of the common dynamical horizon is strongly correlated with the news of the outgoing gravitational radiation, just as expected from \cite{shear-news2020}. Secondly, one can infer the parameters of the final black hole and the parent black holes from the post-merger shear of the common dynamical horizon. 

The common dynamical horizon lives in the strong field regime of the spacetime whereas the gravitational signals we receive are in the weak field region. The common horizon is highly deformed when formed and its geometry is significantly different from the extraction spheres of gravitational wave signals in the far-away regions. The common horizon during the post-merger phase is thus actively absorbing gravitational radiation, quantified by its shear, as it loses hair and settles down to a Kerr isolated horizon. We analyze the shear in the light of the correlation conjecture and find interesting results. 

First, the shear is found to be tightly correlated with the news computed from the numerical simulation. Second, we show that the shear has quasi-normal behaviour and its evolution in the post-merger phase is dominantly described by the fundamental QNM of the Kerr black hole to which the dynamical horizon approaches at late times. The fundamental QNM was recovered with limited accuracy by fitting the shear immediately since the formation of the common horizon to unconstrained damped sinusoids. The spins of the dynamical horizons were also estimated from the obtained best-fit frequencies. Third, we show that the final mass and the spin of the remnants can be also recovered by parameter estimation. Lastly, we also show that the individual masses of the parent black holes can be estimated a good accuracy using a parameter estimation routine that was specially developed and the SEOBNRv4PHM waveform approximant.

These results suggest that the correlation between the strong field and the weak field dynamics of a BBH system carries past the inspiral phase for the systems studied as expected in \cite{shear-news2020}. Also, there is evidence that relations hold for a wide range of mass ratios. Using these results, one can now understand the physics in the strong field, close to the horizons of the system using the gravitational wave information recorded in our detectors. These include, for instance, the amount of energy and angular momentum absorbed by the black holes.

There is room for improvement of the results. The numerical simulations presented here are of limited accuracy due to the lack of computational resources, which we intend to alleviate in the future as we gain access to more resources. Further sources of error can include mode-mixing due to the unconstrained nature of the fits.

\section{Acknowledgments:}
The author thanks Prof. Sukanta Bose, Dr. Anshu Gupta and Prof. Badri Krishnan for their continued encouragement. This work is funded by Shyama Prasad Mukherjee Fellowship (CSIR) and in part by the Department of Atomic Energy, Government of India, under project no. RTI4001. The numerical simulations and other computations were performed on the high-performance supercomputers Perseus and Pegasus at IUCAA and Sonic at ICTS.

\bibliographystyle{apsrev4-1}

\bibliography{references2.bib}
\end{document}